\documentclass[conference]{IEEEtran}
\IEEEoverridecommandlockouts
\usepackage{cite}
\usepackage{amsmath,amssymb,amsfonts}
\usepackage{algorithmic}
\usepackage{algorithm}
\usepackage{graphicx}
\usepackage{textcomp}
\usepackage{xcolor}
\usepackage{multirow}
\usepackage{adjustbox}
\usepackage{subfig}
\usepackage{orcidlink}

\def\BibTeX{{\rm B\kern-.05em{\sc i\kern-.025em b}\kern-.08em
    T\kern-.1667em\lower.7ex\hbox{E}\kern-.125emX}}
\begin{document}

\title{Residue Number System (RNS) based\\ Distributed Quantum Addition\\
\thanks{This research used resources of the Oak Ridge Leadership Computing Facility, which is a DOE Office of Science User Facility supported under Contract DE-AC05-00OR22725.}
}

\author{\IEEEauthorblockN{Bhaskar Gaur} 
\IEEEauthorblockA{\textit{University of Tennessee} \\
Knoxville, TN, USA \\
bgaur@vols.utk.edu}
\and
\IEEEauthorblockN{Travis S. Humble} 
\IEEEauthorblockA{\textit{Oak Ridge National Laboratory} \\
Oak Ridge, TN, USA \\
humblets@ornl.gov}
\and
\IEEEauthorblockN{Himanshu Thapliyal} 
\IEEEauthorblockA{\textit{University of Tennessee} \\
	Knoxville, TN, USA \\
	hthapliyal@utk.edu}
}

\maketitle

\begin{abstract}
Quantum Arithmetic faces limitations such as noise and resource constraints in the current Noisy Intermediate Scale Quantum (NISQ) era quantum computers. We propose using Distributed Quantum Computing (DQC) to overcome these limitations by substituting a higher depth quantum addition circuit with Residue Number System (RNS) based quantum modulo adders. The RNS-based distributed quantum addition circuits possess lower depth and are distributed across multiple quantum computers/jobs, resulting in higher noise resilience. We propose the Quantum Superior Modulo Addition based on RNS Tool (QSMART), which can generate RNS sets of quantum adders based on multiple factors such as depth, range, and efficiency. We also propose a novel design of Quantum Diminished-1 Modulo (2\textsuperscript{n} + 1) Adder (QDMA), which forms a crucial part of RNS-based distributed quantum addition and the QSMART tool. We demonstrate the higher noise resilience of the Residue Number System (RNS) based distributed quantum addition by conducting simulations modeling Quantinuum's H1 ion trap-based quantum computer. Our simulations demonstrate that RNS-based distributed quantum addition has 11.36\% to 133.15\% higher output probability over 6-bit to 10-bit non-distributed quantum full adders, indicating higher noise fidelity. Furthermore, we present a scalable way of achieving distributed quantum addition higher than limited otherwise by the 20-qubit range of Quantinuum H1.
\end{abstract}


\section{Introduction}

Quantum computing is an emerging field that promises considerable advantage over classical computing owing to quantum algorithms such as Shor’s
algorithm, quantum phase estimation, amplitude amplification, variational quantum algorithms,
quantum approximate optimization algorithms, HHL algorithm, and quantum convolution neural networks (QCNN) \cite{shor1994algorithms, duan2020survey, hur2022quantum, lee2019hybrid}. The quantum algorithms rely on quantum arithmetic circuits for representation and optimization. Quantum adders play an essential role among quantum arithmetic circuits by forming building blocks for quantum multiplication, quantum image processing, and quantum cryptanalysis \cite{munoz2018quantum, mohan2016residue, roetteler2017quantum, putranto2022another}. Hence, optimizing quantum adders and increasing their noise resilience is a critical objective.

Recent advancements have been made in quantum computing, particularly by ion trap quantum computers, to achieve higher gate fidelity. Although more reliable, ion trap quantum computers offer fewer qubits than their superconducting counterparts, resulting in quantum circuit size limitations. Also, noise accumulation with increasing depth of quantum circuits results in lower output probability, limiting the scaling potential.
Distributed Quantum Computing (DQC) is an emerging computing paradigm that builds upon the existing individual quantum systems to significantly increase computing power and help scale quantum algorithms to larger sizes \cite{tang2024distributed, barral2024review}. 

This work aims to achieve scalability for quantum adders using a hybrid distributed quantum computing scheme. We use classical computing to distribute quantum addition across multiple quantum circuits using Residue Number Systems (RNS). RNS represents numbers in the form of residues or remainders left after division with a set of relatively prime moduli. The representation is unique as long as the numbers lie within the range of the moduli product. Each of the residues can be added individually without any carry dependency upon other residues in the RNS set. This provides the capability to distribute the addition for residues across multiple quantum computers or jobs. Apart from addition, RNS can also accomplish closure for arithmetic operations such as subtraction and multiplication.

The choice of moduli selection for RNS is an active research area in classical computing. We select a set of three moduli: 2\textsuperscript{n}-1, 2\textsuperscript{n}, and 2\textsuperscript{n}+1 to construct RNS sets in this work. The addition achieved by this RNS set can nearly substitute the range achievable by a quantum full adder of input size '3n' \cite{mohan2016residue}. These moduli can help construct scalable RNS sets with optimized quantum addition implementations that use lower quantum resources.
While designs exist for quantum modulo adders for the moduli (2\textsuperscript{n}-1) and 2\textsuperscript{n}, the quantum modulo (2\textsuperscript{n}+1) adder lacked an implementation optimized specifically for it. Hence, we propose a Quantum Diminished-1 Modulo (2\textsuperscript{n}+1) Adder (QDMA) that can perform (A+B) modulo (2\textsuperscript{n}+1) addition efficiently.

Ultimately, we propose a novel Quantum Superior Modulo Addition based on RNS Tool (QSMART) that automates the moduli selection and generation of RNS based quantum modulo adder circuits for a given range of inputs. The QSMART Tool's approach of using Residue Number System (RNS) for quantum addition: i) splits the quantum addition operation in multiple circuits independently executable in a Distributed Quantum Computing (DQC) system, ii) reduces the overall depth, and iii) achieves higher output probability resulting from superior noise resilience. The proposed QSMART Tool can help achieve higher noise resilience in NISQ computing, and scale quantum addition in NISQ as well as Fault Tolerant Quantum (FTQ) era quantum computers.
We list the contributions of this work below:
\begin{itemize}
	\item We propose Quantum Diminished-1 Modulo (2\textsuperscript{n}+1) Adder (QDMA) that addresses a crucial gap in the creation of Residue Number System (RNS) based quantum addition.
	\item We propose the QSMART Tool, which generates RNS-based quantum modulo adders, leveraging Distributed Quantum Computing (DQC) to execute quantum addition independently.
	\item We demonstrate that the proposed QSMART Tool's reduction in overall depth translates into higher noise resilience by conducting simulations that model the system noise in Quantinuum's H1 ion trap-based quantum computer on PyTKET \cite{Sivarajah_2021}.
\end{itemize}

This paper is organized as follows: Section \ref{Background} provides the background information and relevant works utilized in this paper. Section \ref{proposed adder} proposes the novel Quantum Diminished-1 Modulo (2\textsuperscript{n}+1) Adder (QDMA). Section \ref{proposed qsmart tool} explains the QSMART Tool and its algorithm. Section \ref{results} shows the Quantinuum H1-based simulation results for quantum modulo adders and the noise-resilience of QSMART Tool. Section \ref{discussion}  discusses the resource and output probability comparison between the quantum full adder and distributed quantum addition. Finally, Section \ref{conclusion} concludes this work.

\section{Background}
\label{Background}

\subsection{Quantum Modulo Addition}
Modulo, or modular addition, is an addition operation performed within a modulo arithmetic system in which the output does not exceed the modulo. Equation \ref{equation:ModAddition} shows the modulo addition for a given modulo 'M', where the sum of inputs A and B is divided by the moduli M, and a reminder is provided as a result. 

\begingroup
\setlength\abovedisplayskip{0pt}
\begin{equation} \label{equation:ModAddition}
	\begin{aligned} 				  
		&Modulo \,Sum=\begin{cases}
			A + B, & \text{if}\;{(A+B)}<{M} \\
			0, & \text{if}\;{(A+B)}={M} \\
			(A + B)mod(M), & \text{if}\;{(A+B)}>{M}
		\end{cases} \\[10pt]
		&\text{where }\;0\leq{A,B,Modulo \,Sum}<{M} \\
	\end{aligned}
\end{equation}
\endgroup

\begin{figure}[!t]
	\centering
	\subfloat[]{\includegraphics[width=0.13\textwidth]{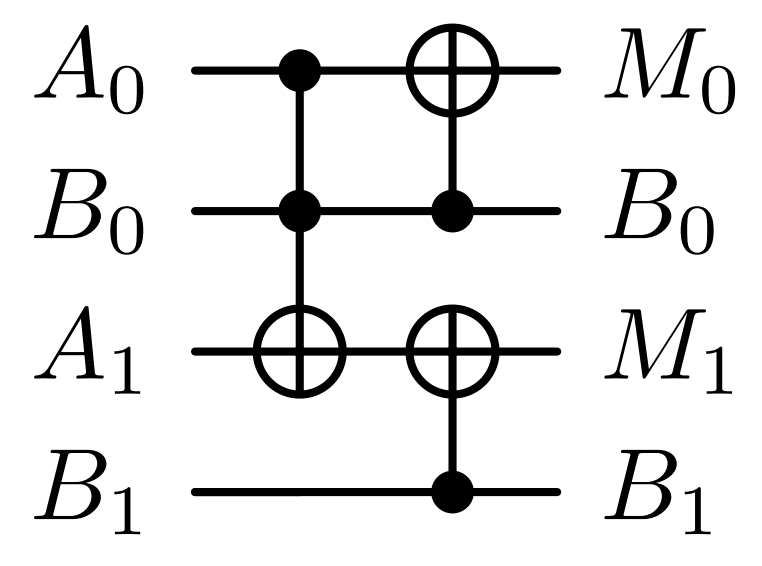}%
	}
	\hfil
	\subfloat[]{\includegraphics[width=0.33\textwidth]{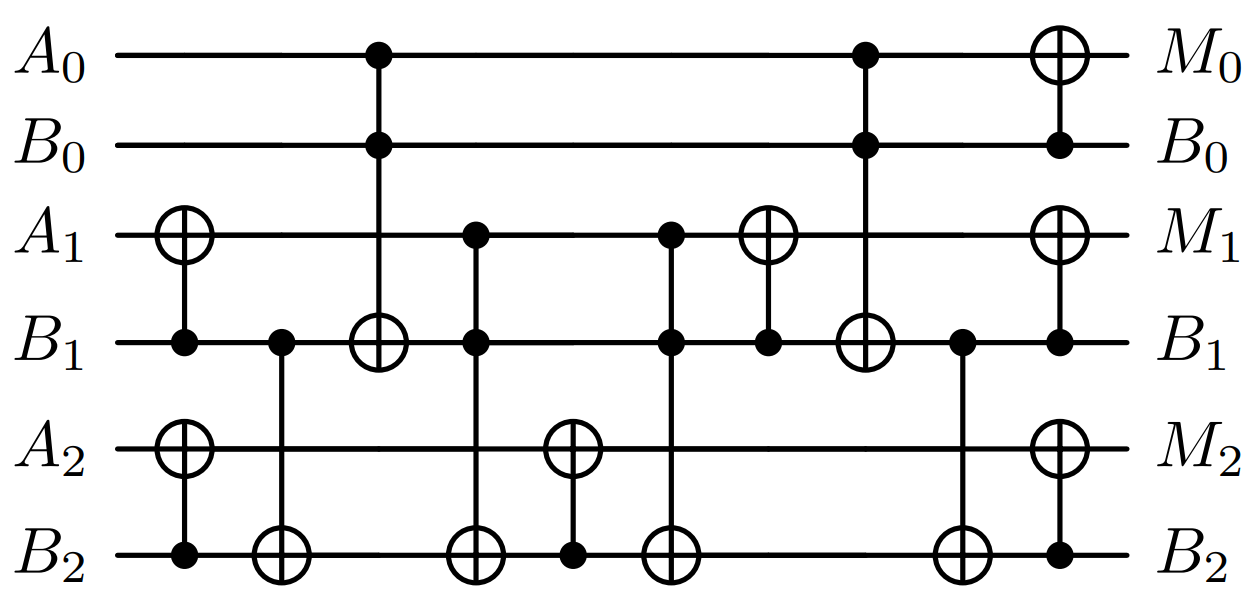}%
	}
	\caption{Quantum Modulo 2\textsuperscript{n} adders used in this work: (a) Mod 4 Adder. (b) Mod 8 Adder. The inputs are A and B. The output M is Modulo Sum, while input B is passed unchanged.}
	\vspace{-0.3cm}
	\label{fig:mod48}
\end{figure}

\begin{figure*}[!t]
	\centering
	\subfloat[]{\includegraphics[width=0.61\textwidth]{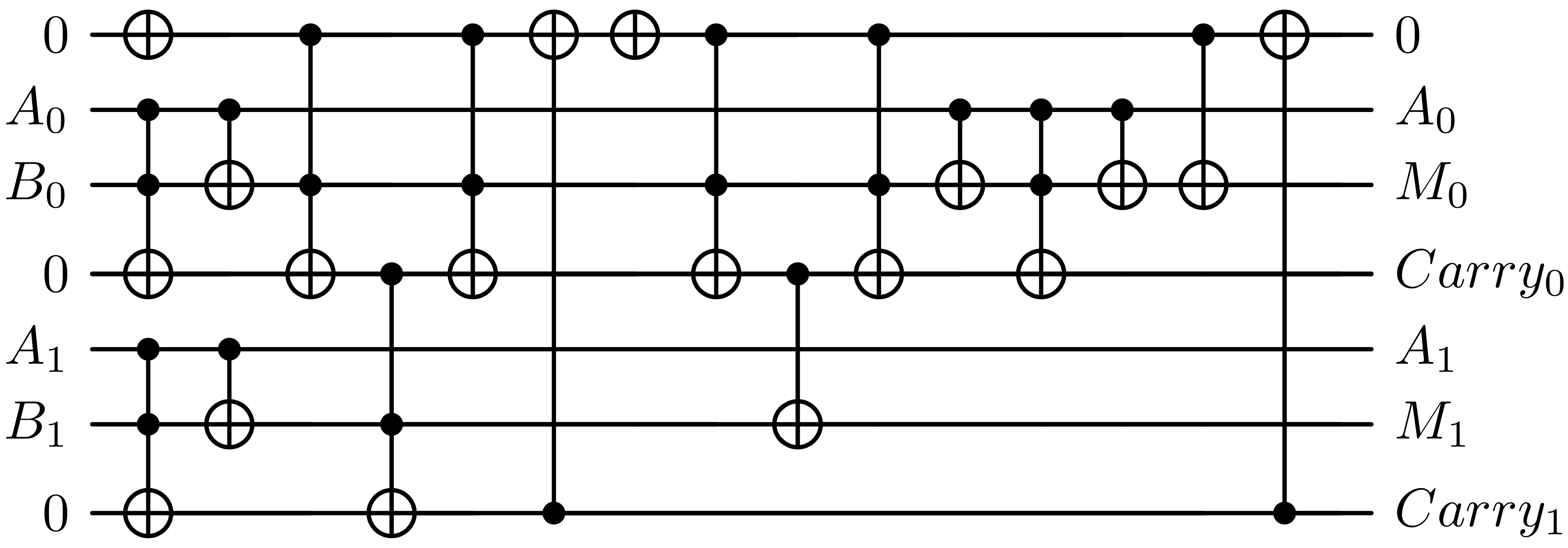}%
	}
	\hfil
	\subfloat[]{\includegraphics[width=0.32\textwidth]{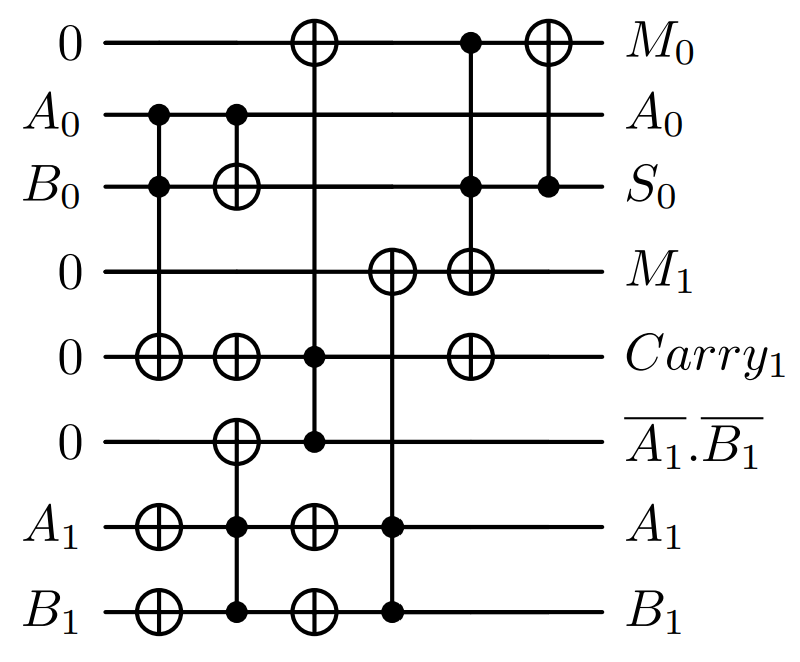}%
	}
	\caption{Comparison of Quantum Modulo 3 adders used in this work: (a) Quantum Modulo (2\textsuperscript{n} - 1) adder for n = 2 configuration \cite{kim2021quantum}. (b) Quantum Modulo (2\textsuperscript{n} + 1) adder for n = 1 configuration. The inputs are A (A\textsubscript{0}:A\textsubscript{1}) and B (B\textsubscript{0}:B\textsubscript{1}), while the output is Modulo Sum M (M\textsubscript{0}:M\textsubscript{1}). Input A is passed without modification. Quantum Modulo (2\textsuperscript{n} + 1) adder's Toffoli count and CNOT count are lower by 3 and 6 respectively. Also, its Toffoli depth and CNOT depth are lower by 3 and 5 respectively, showcasing its better design.}
	\vspace{-0.3cm}
	\label{fig:mod3}
\end{figure*}

Quantum modulo adders are quantum arithmetic circuits optimized especially for modulo addition. They help create quantum modulo circuits of higher complexity, such as subtraction and multiplication, used in applications such as quantum cryptanalysis, quantum image processing, and securing communication \cite{roetteler2017quantum, putranto2022another}. The prior work on quantum modulo adders for generic modulo M has a high overhead regarding quantum gates and depth, limiting their usefulness on the quantum computers available today. The quantum modulo M adders proposed by Vedral et al. and Beauregard et al. have five quantum full adder stages \cite{vedral1996quantum, beauregard2003quantum}. Van Meter and Itoh proposed a more efficient modulo adder with three quantum full adder stages \cite{rodney2005}. In this work, we focus on the moduli that are powers of 2 as: i) they are efficient as they represent the maximum numbers possible in n-bits, and ii) they require fewer quantum resources than quantum modulo M adders. Modulo 2\textsuperscript{n} adders require at most one quantum full adder stage, as any addition without carryout is modulo 2\textsuperscript{n} addition. This is evident from Figure \ref{fig:mod48} in which modulo 8 adder is optimized version of a quantum full adder with 4-qubit output, and modulo 4 adder uses only two CNOT and a Toffoli gates. Modulo (2\textsuperscript{n}-1) adder by Kim et al., on the other hand, requires up to two full adder stages \cite{kim2021quantum}. We propose modulo (2\textsuperscript{n}+1) adder in this work that needs one full and a half adder stage. Figure \ref{fig:mod3} provides a comparison between modulo 3 adders achieved by Kim et al.'s modulo (2\textsuperscript{n}-1) adder and the modulo (2\textsuperscript{n}+1) adder proposed in this work. Ultimately, these three quantum modulo adders help us achieve RNS based arithmetic in the following Section.

\begin{figure}[h]
	\centering
	\includegraphics[scale=0.29]{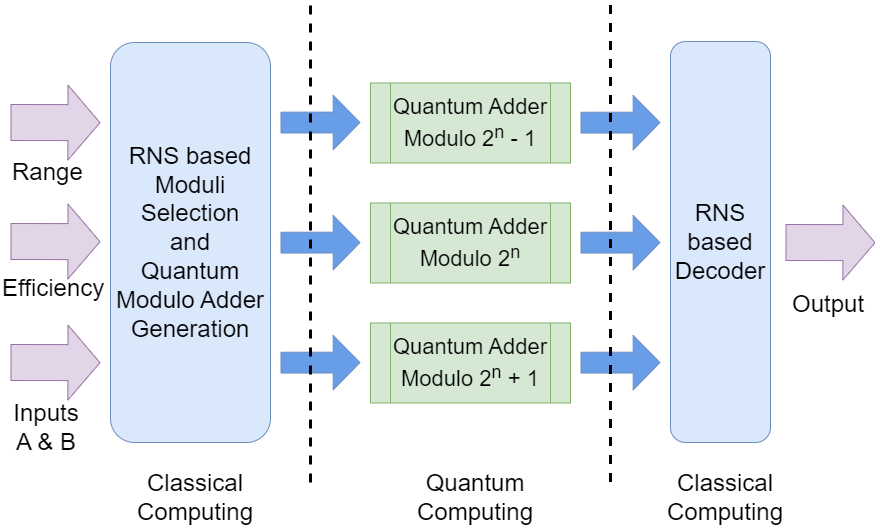}
	\caption{Proposed  Quantum Superior Modulo Addition based on RNS Tool (QSMART) uses a hybrid classical-quantum flow to achieve Distributed Quantum Computing (DQC) for quantum addition. QSMART accepts the range of addition, planned efficiency of representation, and inputs to deliver multiple independent Residue Number System (RNS) based quantum modulo adders that can be distributed across multiple quantum computers or jobs.}
	\label{fig:qsmart}
\end{figure}

\subsection{Residue Number System}

The Residue Number System (RNS) is a numerical system where integers are represented by their residues taken with respect to a set of relatively prime integers called moduli. RNS allows for fast arithmetic operations, particularly addition, subtraction, and multiplication, due to parallel and autonomous processing of residues without carry dependency \cite{mohan2016residue}. RNS can be scaled to higher ranges efficiently by carefully selecting moduli that are large enough to cover the desired range while ensuring they remain relatively prime. Commonly used moduli in RNS include small primes and powers of two. The modulo set used in this work is (2\textsuperscript{n}-1, 2\textsuperscript{n}, 2\textsuperscript{n}+1), that can represent conventional numbers almost up to 2\textsuperscript{3n}. Equation \ref{equation:Range} establishes the Range of the RNS system as the product of the moduli set, where M represents the moduli count, and the moduli set is {R\textsubscript{0}, R\textsubscript{1}..... R\textsubscript{M}}. We define Efficiency of the RNS set, as the ratio of Range and the highest intended number (K), shown in Equation \ref{equation:Efficiency}. Efficiency can help compare and choose among various RNS sets.

\begin{equation} \label{equation:Range}
	\begin{aligned}
		&Range = \prod_{1}^{M} R \\
	\end{aligned}
\end{equation}
\begin{equation} \label{equation:Efficiency}
	\begin{aligned}
		&Efficiency = \frac{Range}{K}
		&\text{where } Range \le K
	\end{aligned}
\end{equation}

In classical computing, RNS finds applications in various fields, such as digital image processing, digital signal processing, cryptography, neural systems, and fault-tolerant computing systems \cite{mohan2016residue}. In quantum computing, RNS can facilitate larger and faster arithmetic operations, overcoming limitations such as qubit count and coherence time of quantum computers. Redundancy in RNS introduces additional control residues to detect and correct errors \cite{kalmykov2022error}. By expanding the code combination with redundant residues, RNS can improve the reliability of quantum computers.

\subsection{Distributed Quantum Computing}

Distributed Quantum Computing (DQC) is an emerging paradigm of performing quantum computations across multiple interconnected quantum computing devices or nodes, also known as Quantum Processing Units (QPU). The goal is to leverage the collective power of multiple QPU's to solve complex computational problems that cannot be efficiently solved by a single QPU. DQC aims to overcome the limitations of individual quantum devices, such as limited qubit counts and high error rates, and to enable the scaling of quantum computations to larger problem sizes\cite{barral2024review}.

The pursuit of DQC involves various approaches, primarily categorized into hardware and software solutions. Hardware solutions, like quantum teleportation enabled using gate or qubit teleportation, necessitate efficient quantum networking schemes for the transmission of quantum states across QPUs, and are a vibrant area of research. On the other hand, software approaches, such as circuit cutting, distribute parts of circuits among QPUs without the need for a fully realized quantum network. However, circuit cutting faces significant challenges, including the exponential increase in cost with the entanglement between parts, the management of different noise profiles, and the reconstruction of output \cite{tang2024distributed}.

Figure \ref{fig:qsmart} shows our approach of using a Residue Number System (RNS) based Distributed Quantum Computing (DQC), that converts quantum addition into independent quantum modulo addition circuits across multiple QPUs or multiple jobs on a single QPU. The QSMART Tool does not need complex networking infrastructure, like in hardware-based DQC approaches. Also, unlike circuit cutting, there is no dependency among different quantum modulo addition circuits of initializing the input of one circuit based on the output of another circuit.

\vspace{2cm}


\section{Proposed Quantum Diminished-1 \\Modulo (2\textsuperscript{n} + 1) Adder}
\label{proposed adder}
In order to achieve modulo (2\textsuperscript{n} + 1) addition, we utilize Algorithm \ref{alg:1} to construct the Quantum Diminished-1 Modulo (2\textsuperscript{n} + 1) Adder (QDMA). The two inputs, A = a\textsubscript{n} a\textsubscript{n-1} ... a\textsubscript{0} and B = b\textsubscript{n} b\textsubscript{n-1} ... b\textsubscript{0}, are of size (n+1) qubits, while their Sum is an (n+2) qubit integer Sum = A + B = S\textsubscript{n+1} S\textsubscript{n} ... S\textsubscript{0}. Algorithm \ref{alg:1} achieves optimized modulo (2\textsuperscript{n} + 1) addition using different components of the Sum.

In order to better handle the complexity introduced for handling 0 in modulo (2\textsuperscript{n} + 1) arithmetic, we utilize diminished-1 representation. In this approach, we subtract one from the numbers in the range (1, 2\textsuperscript{n}), while 0 is represented by 2\textsuperscript{n}. The Most Significant Bit (MSB) of the resulting diminished-1 number is true only when representing 0, helping achieve a more straightforward identification and handling of 0.

\begin{figure}[h]
	\centering
	\includegraphics[scale=0.4]{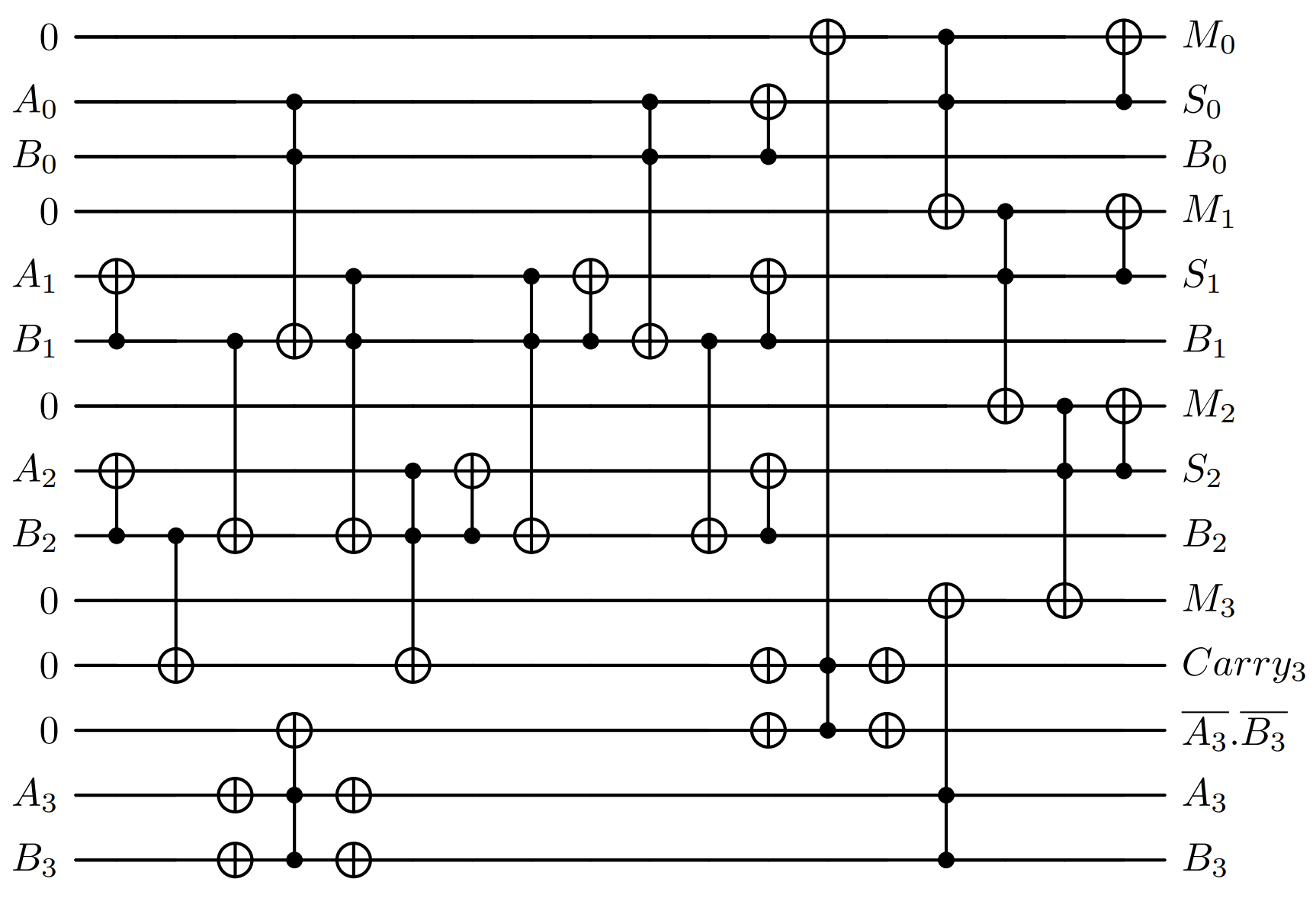}
	\caption{Proposed Quantum Modulo (2\textsuperscript{n} + 1) Adder for n = 3 configuration yielding Quantum Modulo 9 Adder. The inputs are A (A\textsubscript{0}:A\textsubscript{3}) and B (B\textsubscript{0}:B\textsubscript{3}), while the output is Modulo Sum M (M\textsubscript{0}:M\textsubscript{3}). Input B (B\textsubscript{0}:B\textsubscript{3}) and MSB of A (A\textsubscript{3}) are passed unchanged. The Sum (S\textsubscript{0}:S\textsubscript{2}:Carry\textsubscript{3}) represents the sum of (n-1) qubits of inputs A and B. As shown in Algorithm \ref{alg:1}, first the Sum is calculated, with the Carry\textsubscript{3} used to compute $\overline{A\textsubscript{3}}.\overline{B\textsubscript{3}}.\overline{Carry\textsubscript{3}}$.  Finally, quantum half adder stage calculates the Modulo Sum M after adding (A\textsubscript{3}.B\textsubscript{3}, S\textsubscript{2} ... S\textsubscript{0}) + $\overline{A\textsubscript{3}}.\overline{B\textsubscript{3}}.\overline{Carry\textsubscript{3}}$}
	\label{fig:mod9}
\end{figure}

The algorithm \label{alg:1} reflects the advantage of using the diminished-1 approach. The addition in Step 1 saves resources by having to add only (n-1) qubits, as it handles the (1, 2\textsuperscript{n}). Step 2 calculates the partial product $\overline{A\textsubscript{n}}.\overline{B\textsubscript{n}}.\overline{Carry\textsubscript{n}}$ which is True only if both input's MSB and Step1's Carry are zero. To create the three input quantum NAND gate, we utilize two Toffoli gates and six NOT gates, one each for the inputs and later for the reversibility. The resulting partial product is then placed at the Least Significant Bit (LSB) of the QDMA, initialized with \text{$|0$}⟩. Step 3 utilizes a quantum half adder to ultimately add the previous step's partial product. The term A\textsubscript{n}.B\textsubscript{n} in Step 3 helps achieve the addition when both inputs are zero. Figure \ref{fig:mod9} shows the proposed QDMA for modulo 9, achieved when used in n=3 configuration.

\begin{algorithm}
	\renewcommand{\algorithmicrequire}{\textbf{Input:}}
	\caption{: Calculation of $(A + B) mod (2^n + 1)$}\label{alg:1}
	\begin{algorithmic}
		\REQUIRE Inputs A, B are in diminished-1 format. $n \geq 1$
		\ENSURE $0\leq{A,B} < {2^n+1}$
		\STATE 1. Add the inputs except their most significant bits (MSB). $Sum = A\textsubscript{n-1}A\textsubscript{n-2} ... A\textsubscript{0} + B\textsubscript{n-1}B\textsubscript{n-2} ... B\textsubscript{0} = Carry\textsubscript{n} S\textsubscript{n-1} ... S\textsubscript{0}$
		\STATE 2. Using the Carry from Step 2, compute: \\ \centerline{$\overline{A\textsubscript{n}}.\overline{B\textsubscript{n}}.\overline{Carry\textsubscript{n}}$}
		\STATE 3. $Modulo Sum = (A\textsubscript{n}.B\textsubscript{n}, S\textsubscript{n-1} ... S\textsubscript{0}) + \overline{A\textsubscript{n}}.\overline{B\textsubscript{n}}.\overline{Carry\textsubscript{n}}$
	\end{algorithmic}
\end{algorithm}

As the output or Modulo Sum M cannot exceed the modulus, its size is limited to (n+1)-bits and is represented as M = M\textsubscript{n} ... M\textsubscript{0}. We use a quantum full adder called TPL13 without input carry for: i) to save quantum resources, and ii) its double inverted structure reduces the gap between qubit operations, preventing coherence time issues and reducing noise \cite{thapliyal2013design, bgaur2023noise}.

\section{Proposed QSMART Tool}
\label{proposed qsmart tool}
In this Section, we propose the QSMART Tool that generates Residue Number System (RNS) based quantum modulo adders given input constraints such as the range of numbers to be represented (0, K-1) and the desired minimum efficiency of the RNS set (E). As shown in Algorithm \ref{alg:2}, we start with a default moduli count (C) three, as it corresponds to our starting RNS moduli set of (2\textsuperscript{n}-1, 2\textsuperscript{n}, 2\textsuperscript{n}+1). The advantage of this set is the scaling it provides in representing the range of numbers up to 2\textsuperscript{3n}. For instance, n = 2 yields a moduli set (3, 4, 5) that can represent a range up to (3x4x5) 60, close to (2\textsuperscript{3x2}) 64. Since the last four numbers cannot be represented, the efficiency of this RNS moduli set is 93.75\%. However, the tool can be pushed to limits by choosing E = 1.0, leading to the choice of a moduli set that yields higher ranges.

\begin{algorithm}
	\renewcommand{\algorithmicrequire}{\textbf{Inputs:}}
	\renewcommand{\algorithmicensure}{\textbf{Output:}}
	\caption{: QSMART Tool}\label{alg:2}
	\begin{algorithmic}
		\REQUIRE 1. Number K to be represented (K $\ge 50$).\\ 2. Moduli count C (default value 3).\\ 3. RNS efficiency E (default value 0.9).\\ 4. Quantum modulo adder set: (2\textsuperscript{n}-1, 2\textsuperscript{n}, 2\textsuperscript{n}+1)
		\ENSURE RNS set (R\textsubscript{1}, R\textsubscript{2}, R\textsubscript{3}) to be created.
		\STATE 1. If K $\approx$ 2\textsuperscript{3h}, where h $\geq$ 2. Select R where:
		\\ \centerline{R = (2\textsuperscript{h}-1, 2\textsuperscript{h}, 2\textsuperscript{h}+1)}
		\STATE 2. Return R if $R_{1} \ast R_{2} \ast R_{3} \geq E \ast K$ . Else go to Step 3.
		\STATE 3. Select lowest moduli from the moduli set such that: 
		\\ \centerline{$\prod_{1}^{C} R \ge E \ast K $}
		\STATE 4. If cannot satisfy Step 1, increment C and repeat Step 1.
		\STATE 5. Determine the Toffoli depth for quantum adders for each moduli in set R.
		\STATE 6. For moduli with maximum Toffoli depth, check alternate moduli with lower Toffoli depth that meets constraint E.
		\STATE 7. Return R once cannot optimize any further in Step 3.
	\end{algorithmic}
\end{algorithm}

For the range of numbers that do not correspond to 2\textsuperscript{3n}, we select other moduli created using one of the moduli from 2\textsuperscript{n}-1, 2\textsuperscript{n} or 2\textsuperscript{n}+1. For instance, when representing (2\textsuperscript{8}) 256, we choose the set (5, 8, 9) as it can provide us with the desired range using relatively prime moduli although derived using different values of 'n.' Moduli (2\textsuperscript{2}+1) 5, (2\textsuperscript{3}) 8, and (2\textsuperscript{3}+1) 9 can represent a range up to (5x8x9) 360, that is much higher than the needed 256. Although we could have chosen moduli 7 instead of 9, as the set (5x8x7) would have fulfilled all requirements, we avoided it as the depth of quantum modulo 7 adder is higher than quantum modulo 9 adder by three Toffoli gates.

\section{Results}
\label{results}
We conduct simulations using an emulator provided by Quantinuum to simulate their 20-qubit H1 quantum computer. We perform these simulations for TPL13-based quantum full adder from 6-bit to 10-bit output and quantum modulo adders between 2 and 9 for the moduli set (2\textsuperscript{n}-1, 2\textsuperscript{n}, 2\textsuperscript{n}+1). We initialize circuits for these designs in \text{$|0$}⟩ and \text{$|1$}⟩ basis states on the PyTKET platform. We simulate quantum modulo adders with a frequency of a hundred shots, whereas the TPL13-based quantum full adders are simulated with two hundred shots. To calculate the output probability, we take the ratio of the correct output's frequency with total frequency and take an average across all the input combinations.

\subsection{Quantum Modulo Adders}
Table \ref{table:moduli} shows the output probability and quantum resource usage for the quantum modulo adders. These include moduli (2\textsuperscript{n}-1, 2\textsuperscript{n}, 2\textsuperscript{n}+1), which helps perform RNS based quantum addition in this work. The primary observation is the direct correlation of output probability to Toffoli depth. The output probability gradually falls as the Toffoli depth increases from modulo 2 to modulo 9. However, this effect is seen at different rates among adders of the three types.

\begin{table}[h]
	\centering
	\caption{Comparison of Quantum Moduli Adders based on \\Quantum Resource Usage and Output Probability.}
	\label{table:moduli}
	\begin{adjustbox}{max width=\linewidth}
	\begin{tabular}{|cl|c|cc|cc|c|}
		\hline
		\multicolumn{2}{|c|}{\textbf{Modulo Adder}} & \multirow{2}{*}{\textbf{Qubit}} & \multicolumn{2}{c|}{\textbf{Depth}} & \multicolumn{2}{c|}{\textbf{Count}} & \multirow{2}{*}{\textbf{\begin{tabular}[c]{@{}c@{}}Output\\ Prob.\end{tabular}}} \\ \cline{1-2} \cline{4-7}
		\multicolumn{1}{|c|}{\textbf{Mod}} & \multicolumn{1}{c|}{\textbf{Type}} &  & \multicolumn{1}{c|}{\textbf{Toffoli}} & \textbf{CNOT} & \multicolumn{1}{c|}{\textbf{Toffoli}} & \textbf{CNOT} &  \\ \hline
		\multicolumn{1}{|c|}{\textbf{2}} & 2\textsuperscript{n} & 2 & \multicolumn{1}{c|}{0} & 1 & \multicolumn{1}{c|}{0} & 1 & 0.995 \\ \hline
		\multicolumn{1}{|c|}{\textbf{3}} & 2\textsuperscript{n}-1 & 7 & \multicolumn{1}{c|}{6} & 7 & \multicolumn{1}{c|}{8} & 8 & 0.912 \\ \hline
		\multicolumn{1}{|c|}{\textbf{3}} & 2\textsuperscript{n}+1 & 8 & \multicolumn{1}{c|}{4} & 2 & \multicolumn{1}{c|}{5} & 2 & 0.964 \\ \hline
		\multicolumn{1}{|c|}{\textbf{4}} & 2\textsuperscript{n} & 4 & \multicolumn{1}{c|}{1} & 1 & \multicolumn{1}{c|}{1} & 2 & 0.985 \\ \hline
		\multicolumn{1}{|c|}{\textbf{5}} & 2\textsuperscript{n}+1 & 11 & \multicolumn{1}{c|}{6} & 5 & \multicolumn{1}{c|}{8} & 7 & 0.931 \\ \hline
		\multicolumn{1}{|c|}{\textbf{7}} & 2\textsuperscript{n}-1 & 10 & \multicolumn{1}{c|}{12} & 10 & \multicolumn{1}{c|}{14} & 12 & 0.865 \\ \hline
		\multicolumn{1}{|c|}{\textbf{8}} & 2\textsuperscript{n} & 6 & \multicolumn{1}{c|}{3} & 4 & \multicolumn{1}{c|}{3} & 6 & 0.966 \\ \hline
		\multicolumn{1}{|c|}{\textbf{9}} & 2\textsuperscript{n}+1 & 14 & \multicolumn{1}{c|}{9} & 7 & \multicolumn{1}{c|}{11} & 13 & 0.893 \\ \hline
	\end{tabular}
\end{adjustbox}
\end{table}

Quantum modulo 2\textsuperscript{n} adders have the highest output probability among all the moduli because of their optimal design. For a modulo 2 quantum adder, only a single CNOT gate is needed. As evident from Figure \ref{fig:mod48}, for a modulo 4 quantum adder only one Toffoli with two CNOT gates are needed, and for the modulo 8 quantum adder, we utilize the TPL13 quantum full adder by removing the gates and qubits associated with carry generation. Our optimization for modulo 2\textsuperscript{n} conversion reduces Toffoli count and depth from five to three, a reduction of 40\%. Similarly, the CNOT gate count reduces from ten to six, a drop of 40\%, and the CNOT depth reduces from seven to four, a drop of 42.85\%.

The Quantum Diminished-1 Modulo (2\textsuperscript{n} + 1) Adders (QDMA) are the next best performing in Table \ref{table:moduli}. They utilize fewer Toffoli and CNOT gates compared to Quantum Modulo (2\textsuperscript{n} - 1) Adders (QMA) albeit with higher qubit count, leading to lesser depth and higher output probability. This is evident from the quantum modulo 3 adder, depicted in Figure \ref{fig:mod3}, whose QDMA-based design has 33\% lower Toffoli depth and 71.4\% lower CNOT depth compared to the QMA-based design, leading to 5.7\% higher output probability. For n=3, the QDMA-based modulo 9 adder has 25\% lower Toffoli depth and 30\% lower CNOT depth relative to the QMA-based modulo 7 adder, maintaining 3.27\% higher output probability. 

\begin{table*}[!htbp]
	\centering
	\caption{Comparison of TPL13 based Quantum Full Adders with QSMART Tool's RNS based distributed quantum addition.}
	\label{table:comparison}	
	\begin{tabular}{|c|cccc|ccccccc|}
		\hline
		\multirow{3}{*}{\textbf{\begin{tabular}[c]{@{}c@{}}Adder\\ Size\end{tabular}}} & \multicolumn{4}{c|}{\textbf{Non-Distributed Quantum Addition (TPL13)}} & \multicolumn{7}{c|}{\textbf{Distributed Quantum Addition (RNS)}} \\ \cline{2-12} 
		& \multicolumn{1}{c|}{\multirow{2}{*}{\textbf{\begin{tabular}[c]{@{}c@{}}Qubit\\ Count\end{tabular}}}} & \multicolumn{2}{c|}{\textbf{Depth}} & \multirow{2}{*}{\textbf{\begin{tabular}[c]{@{}c@{}}Output Prob.\end{tabular}}} & \multicolumn{1}{c|}{\multirow{2}{*}{\textbf{RNS Set}}} & \multicolumn{1}{c|}{\multirow{2}{*}{\textbf{\begin{tabular}[c]{@{}c@{}}Efficiency\\ (in \%)\end{tabular}}}} & \multicolumn{1}{c|}{\multirow{2}{*}{\textbf{\begin{tabular}[c]{@{}c@{}}Max.\\ Qubits\end{tabular}}}} & \multicolumn{2}{c|}{\textbf{Max. Depth}} & \multicolumn{1}{c|}{\multirow{2}{*}{\textbf{\begin{tabular}[c]{@{}c@{}}Output\\ Prob.\end{tabular}}}} & \multirow{2}{*}{\textbf{\begin{tabular}[c]{@{}c@{}}Gain w.r.t\\ TPL13 (\%)\end{tabular}}} \\ \cline{3-4} \cline{9-10}
		& \multicolumn{1}{c|}{} & \multicolumn{1}{c|}{\textbf{Toffoli}} & \multicolumn{1}{c|}{\textbf{CNOT}} &  & \multicolumn{1}{c|}{} & \multicolumn{1}{c|}{} & \multicolumn{1}{c|}{} & \multicolumn{1}{c|}{\textbf{Toffoli}} & \multicolumn{1}{c|}{\textbf{CNOT}} & \multicolumn{1}{c|}{} &  \\ \hline
		\textbf{6} & \multicolumn{1}{c|}{11} & \multicolumn{1}{c|}{9} & \multicolumn{1}{c|}{13} & 0.836 & \multicolumn{1}{c|}{(3, 4, 5)} & \multicolumn{1}{c|}{93.75} & \multicolumn{1}{c|}{11} & \multicolumn{1}{c|}{6} & \multicolumn{1}{c|}{5} & \multicolumn{1}{c|}{0.931} & 11.36 \\ \hline
		\textbf{7} & \multicolumn{1}{c|}{13} & \multicolumn{1}{c|}{11} & \multicolumn{1}{c|}{16} & 0.702 & \multicolumn{1}{c|}{(4, 5, 9)} & \multicolumn{1}{c|}{100} & \multicolumn{1}{c|}{14} & \multicolumn{1}{c|}{9} & \multicolumn{1}{c|}{7} & \multicolumn{1}{c|}{0.893} & 27.21 \\ \hline
		\textbf{8} & \multicolumn{1}{c|}{15} & \multicolumn{1}{c|}{13} & \multicolumn{1}{c|}{19} & 0.595 & \multicolumn{1}{c|}{(5, 8, 9)} & \multicolumn{1}{c|}{100} & \multicolumn{1}{c|}{14} & \multicolumn{1}{c|}{9} & \multicolumn{1}{c|}{7} & \multicolumn{1}{c|}{0.893} & 50.08 \\ \hline
		\textbf{9} & \multicolumn{1}{c|}{17} & \multicolumn{1}{c|}{15} & \multicolumn{1}{c|}{22} & 0.481 & \multicolumn{1}{c|}{(7, 8, 9)} & \multicolumn{1}{c|}{98.44} & \multicolumn{1}{c|}{14} & \multicolumn{1}{c|}{12} & \multicolumn{1}{c|}{10} & \multicolumn{1}{c|}{0.862} & 79.21 \\ \hline
		\textbf{10} & \multicolumn{1}{c|}{19} & \multicolumn{1}{c|}{17} & \multicolumn{1}{c|}{25} & 0.371 & \multicolumn{1}{c|}{(4, 5, 7, 9)} & \multicolumn{1}{c|}{100} & \multicolumn{1}{c|}{14} & \multicolumn{1}{c|}{12} & \multicolumn{1}{c|}{10} & \multicolumn{1}{c|}{0.865} & 133.15 \\ \hline
		\textbf{11} & \multicolumn{1}{c|}{21} & \multicolumn{1}{c|}{19} & \multicolumn{1}{c|}{28} & N.A. & \multicolumn{1}{c|}{(5, 7, 8, 9)} & \multicolumn{1}{c|}{100} & \multicolumn{1}{c|}{14} & \multicolumn{1}{c|}{12} & \multicolumn{1}{c|}{10} & \multicolumn{1}{c|}{0.865} & N.A. \\ \hline
	\end{tabular}
\end{table*}

\subsection{Noise-Resilience of QSMART Tool}

Table \ref{table:comparison} compares quantum addition attained via TPL13 quantum full adder and the QSMART tool's RNS based distributed quantum addition on Quantinuum H1 quantum computer's noise-based simulations. For each adder size of TPL13, the equivalent RNS set computed by QSMART Tool (with a default efficiency setting of 90\%) is shown. In Table \ref{table:comparison}, RNS efficiency denotes the percentage of the range covered by the RNS moduli set. For RNS, the maximum qubits or depth are the maximum qubit count or depth, respectively, among the quantum modulo adder circuits of the RNS set. Similarly, the output probability of the RNS set is the minimum output probability across the quantum modulo adder circuits comprising the RNS set. Ultimately, we list the percentage gain in output probability achieved by the RNS set over the corresponding TPL13-based adder.

Table \ref{table:comparison} shows a clear trend of rising output probability gain in favor of distributed quantum addition (based on RNS) compared to non-distributed quantum addition (based on TPL13). Compared to the TPL13 adder of 5 qubits input size, the equivalent RNS set (3, 4, 5) has Toffoli depth reduced by three gates, and CNOT depth reduced to less than half, leading to 11.36\% higher output probability. However, as the effective range of the RNS set is 60 compared to 64 of TPL13, the RNS efficiency is 93.75\%. 

\section{Discussion}
\label{discussion}
\begin{figure}[h]
	\centering
	\includegraphics[scale=0.35]{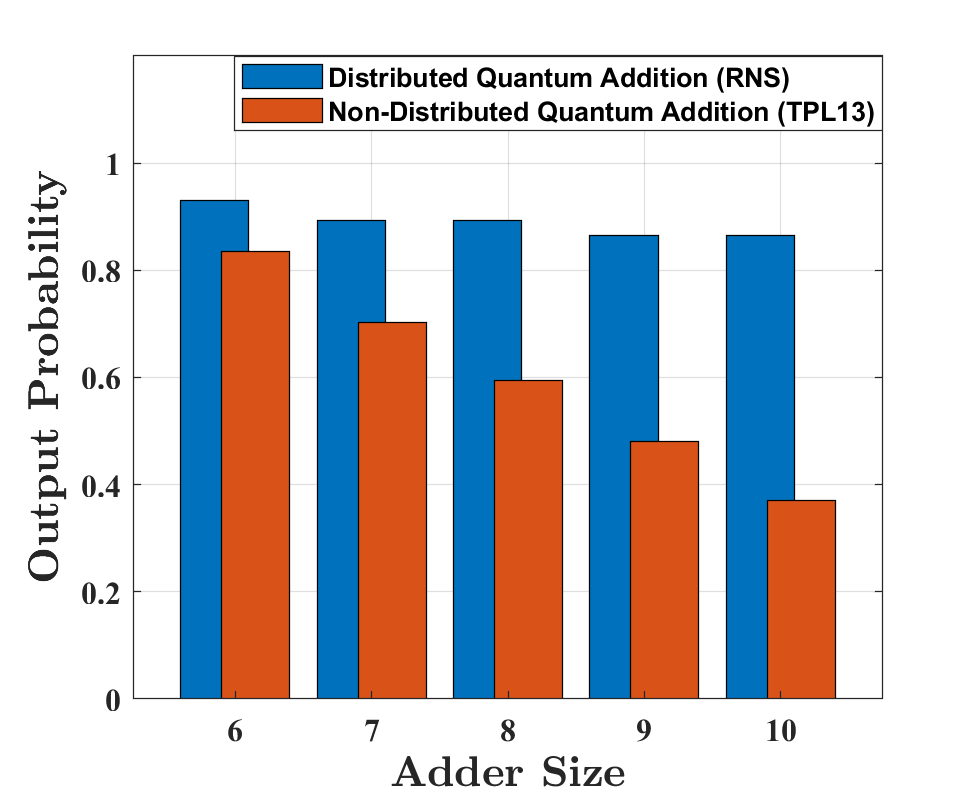}
	\caption{Comparison of Output Probability between TPL13 based quantum full adders and QSMART Tool's RNS based distributed quantum addition for output sizes 6 to 10 qubit.}
	\label{fig:graph}
\end{figure}

As we scale the adder output size from 6 to 10 qubits, the output probability drops from 0.836 to 0.371. However, for the corresponding RNS sets, the output probability declines from 0.931 to 0.865. The lower rate of decline for the RNS set demonstrates its noise resilience on the Quantinuum H1 quantum computer's noise-based simulations. Using a higher efficiency of 1.0 in the QSMART Tool creates an RNS set (3, 5, 8) with twice the effective range than before. Quantum modulo 8 adder consumes higher resources than quantum modulo 4 adder. However, since the quantum modulo 8 adder consumes fewer resources than the other two moduli in the RNS set, there is no change in the maximum qubit count, depth, or output probability. Hence, as shown in Figure \ref{fig:graph}, the advantage of RNS to run three independent quantum jobs, helps Distributed Quantum Computing (DQC) to limit the loss of output probability.

As shown in the last entry of Table \ref{table:comparison}, we can achieve addition equivalent to a quantum full adder of output size 11 or (2\textsuperscript{11}) 2048 by using the RNS set (5, 7, 8, 9). Since this requires 21 qubits, it cannot be performed on Quantinuum's H1 quantum computer. This shows the potential for scaling by QSMART Tool to achieve addition beyond the range of NISQ-era quantum computers. For achieving higher ranges, we can scale the tool with the moduli set (2\textsuperscript{n}-1, 2\textsuperscript{n}, 2\textsuperscript{n}+1) and also use the prime moduli described in Table \ref{table:moduli} to help keep quantum resource consumption in check.

\section{Conclusion}
\label{conclusion}
We conclude that Residue Number Systems (RNS) have great potential to achieve Distributed Quantum Computing (DQC) in the current NISQ era quantum computers by increasing noise-resilience and also in the Fault Tolerant Quantum (FTQ) computers by helping run smaller depth quantum arithmetic circuits independently across machines or jobs. The proposed QSMART Tool has shown potential in achieving these objectives by demonstrating a gain in output probability of 11.36\% to 133.15\% compared to non-distributed quantum addition of output sized six to ten qubits on Quantinuum H1 noise simulator. Furthermore, we show the possibility of achieving quantum addition for larger sizes unachievable by the Quantinuum H1 quantum computer. In the future, quantum multiplication, quantum image, and signal processing tasks could benefit from the speed and parallelism offered by RNS.

\bibliographystyle{IEEEtran}
\bibliography{IEEEabrv, references}

\end{document}